%% file: SHiP-charm.tex
\documentclass[twocolumn,epjc3]{svjour3}          % twocolumn

\RequirePackage[T1]{fontenc}
\smartqed  % flush right qed marks, e.g. at end of proof
\usepackage[utf8]{inputenc}
\RequirePackage{graphicx} 
\usepackage{subfig}
\usepackage{graphicx}
\usepackage{longtable}
\usepackage[abs]{overpic}
\usepackage{multirow}
\usepackage{braket}
\usepackage{lipsum}
\usepackage{comment}
\usepackage{amsmath}
\usepackage{amssymb}
\usepackage{siunitx}
\usepackage{booktabs}
\usepackage{upgreek}

\usepackage{color}   %May be necessary if you want to color links
\usepackage{hyperref}
\usepackage[switch]{lineno}
\usepackage{scalerel,amssymb}

\usepackage{ifthen}  
\newboolean{uprightparticles} 
\setboolean{uprightparticles}{false} %Set true for upright particle symbols 

\RequirePackage{upgreek}
\RequirePackage{graphicx}
\RequirePackage{mathptmx}      % use Times fonts if available on your TeX system
\RequirePackage{flushend}
\RequirePackage[numbers,sort&compress]{natbib}
\usepackage[leftcaption]{sidecap}
\usepackage{graphics}
\pdfoutput=1
\usepackage{comment}
\usepackage{xcolor}
\usepackage{xurl}
\usepackage{float}
\usepackage{testhyphens}
\usepackage{hyphenat}
\usepackage{amsmath}
\usepackage{mathptmx}

\journalname{Eur. Phys. J. C}

\makeatletter

\newcommand{\checknextarg}{\@ifnextchar\bgroup{\nolinebreak\gobblenextarg}{}}
\newcommand{\gobblenextarg}[1]{ \textsuperscript{\nolinebreak\hspace{-4pt}\mbox{\nolinebreak$^,$\nolinebreak\ref{#1}\nolinebreak}\nolinebreak} \@ifnextchar\bgroup{\gobblenextarg}{}}

\begin{document}

\title{Reconstruction of 400 GeV/c proton interactions with the SHiP-charm project}

\titlerunning{SHiP-charm}        % if too long for running head

\author{The SHiP Collaboration$^\text{\normalfont *}$}

\institute{*e-mail: Antonia.Di.Crescenzo@cern.ch \\
*e-mail: Antonio.Iuliano@cern.ch}
%% %simple case: 2 authors, same institution
%% \author{A. Uthor}
%% \author{and A. Nother Author}
%% \affiliation{Institution,\\Address, Country}

% more complex case: 4 authors, 3 institutions, 2 footnotes
%\author[a,b,1]{F. Irst,\note{Corresponding author.}}
%\author[c]{S. Econd,}
%\author[a,2]{T. Hird\note{Also at Some University.}}
%\author[c,2]{and Fourth}

% The "\note" macro will give a warning: "Ignoring empty anchor..."
% you can safely ignore it.

%\affiliation[a]{One University,\\some-street, Country}
%\affiliation[b]{Another University,\\different-address, Country}
%\affiliation[c]{A School for Advanced Studies,\\some-location, Country}

% e-mail addresses: only for the forresponding author
%\emailAdd{Antonia.Di.Crescenzo@cern.ch, Antonio.Iuliano@cern.ch}
\maketitle 

\begin{abstract}
  \noindent
The SHiP-charm project was proposed to measure the associated charm production induced by 400 GeV/c protons in a thick target, including the contribution from cascade production.
An optimisation run was performed in July 2018 at CERN SPS using a hybrid setup. The high resolution of nuclear emulsions acting as vertex detector was complemented by electronic detectors for kinematic measurements and muon identification.
Here we present first results on the analysis of nuclear emulsions exposed in the 2018 run, which prove the capability of reconstructing proton interaction vertices in a harsh environment, where the signal is largely dominated by secondary particles produced in hadronic and electromagnetic showers within the lead target.

\end{abstract}

\input{sections/introduction}

\input{sections/detector_layout}

\input{sections/exposure}
\input{sections/simulation}
\input{sections/event_reconstruction}

\input{sections/vertex_identification}
\input{sections/results}
\input{sections/conclusions}

\input{sections/acknowledgments}

\bibliographystyle{epjc} %
\bibliography{SHiP-charm}%
\input{authorlist_15march2024}
\end{document}

%% file: sections/introduction.tex
\section{Introduction}

The SHiP-charm project \cite{Akmete:2286844} aims at measuring the differential charm production cross
section in a thick target, including the enhancement due to cascade production, which 
has never been measured so far.  Elastic scattering followed by a deep inelastic interaction is the main source of this enhancement~\cite{CASCADE}.
The  accurate  prediction  of  charm  hadroproduction  rates  is  an  essential  ingredient  to establish  the  sensitivity  of  a  high-intensity  proton  beam  dump  experiment  like  SHiP (Search  for  Hidden  Particles)  \cite{Anelli:2015pba}  to new particles produced in charm decays and to make a precise estimation of the tau-neutrino flux at SHiP.
%The knowledge of the associated charm production yield in 400 GeV/c proton interactions is crucial for the SHiP experiment both for Hidden Sector searches and for Neutrino Physics studies. 

An optimization run was performed in July 2018 at the H4 beam line of CERN SPS/North Area, in the same location used for the muon flux measurement~\cite{SHiP:2020hyy}.
A thick target made of lead
interleaved with nuclear emulsions was exposed to a 400 GeV/c proton-beam.
The detector is a hybrid system, combining  Emulsion Cloud Chambers with electronically-read-out detectors, a spectrometer magnet to provide the charge and momentum measurement of charmed-hadron-decay daughters and a muon identification system.

The challenge of the SHiP-charm measurement is two-fold: reconstruct tracks and interaction vertices in a high-density environment and search for rare decays of charmed hadrons. 
The track reconstruction and matching between emulsion and silicon pixel detectors in this optimization run has been described elsewhere \cite{SHiP:2021kxh}.
Here we focus on the identification of interaction vertices, whose success is a prerequisite for subsequent phases
of the analysis.

The work presented in this paper not only contributes to understanding and addressing challenges in track and vertex reconstruction in high-density environments but also establishes the groundwork for broader applications in similar experimental contexts. In particular, our approach is of significant interest to the SND@LHC experiment~\cite{Ahdida:2750060,SNDLHC:2022ihg}, currently collecting data at CERN, which aims to study high-energy neutrinos produced by accelerators for the first time. Furthermore, our work may be of interest to the SHiP@ECN3~\cite{Albanese:2878604}  experiment proposal, which includes a neutrino detector  for conducting high-statistics studies of neutrinos. Both experiments operate in high-density environments, up to 10$^6$ \linebreak tracks/cm$^2$, primarily due to background tracks, all nearly incident at the same angle. This aspect differentiates these experiments from SHiP-charm and introduces additional challenges related to alignment and tracking accuracy, challenges not present in SHiP-charm, where high density is associated with particles produced in secondary interactions and electromagnetic showers.
These connections with current experiments and future proposals underscore the broad relevance and potential applications of the methodology developed in our work.

%In this work we report the developemt

%In the present document we describe the detector layout of the SHiP-charm measurement and the 2018 data taking. 

%% file: sections/detector_layout.tex
\section{Detector layout}

\begin{figure*}[htbp]
\centering
\includegraphics[scale=0.55]{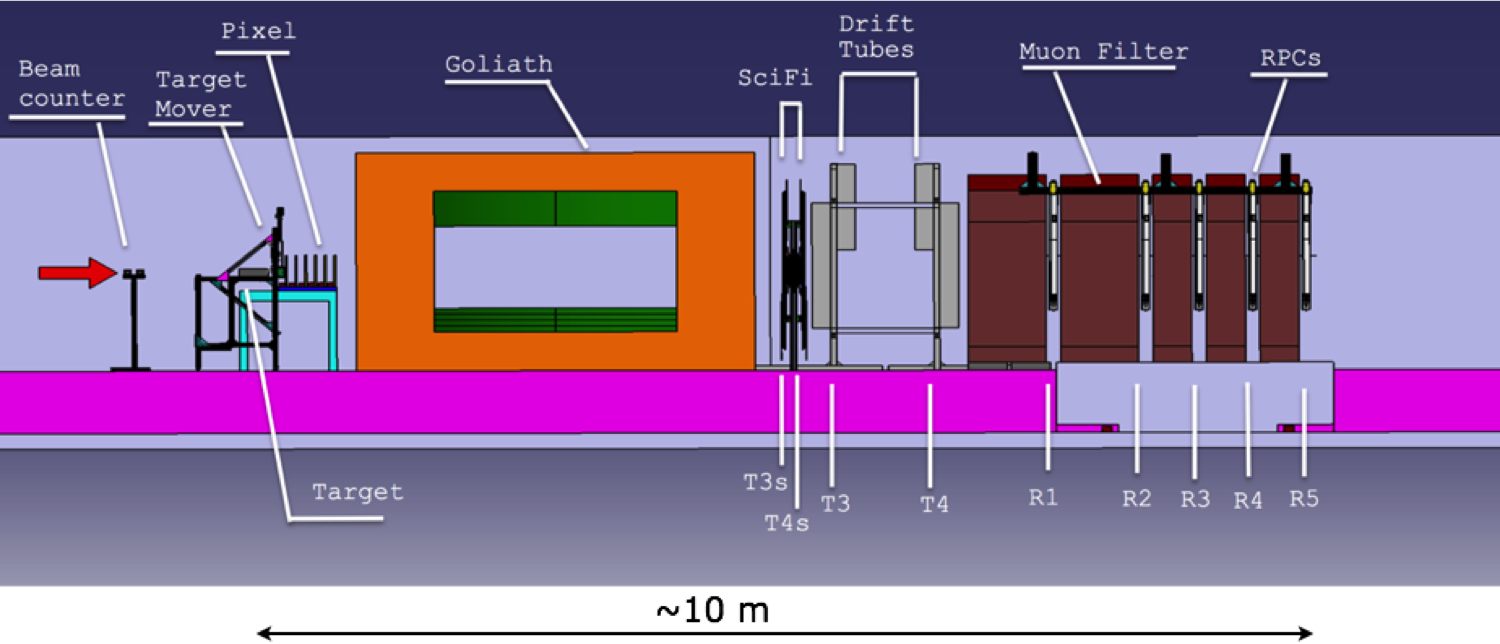}
\caption{Lateral view of the experimental apparatus for the charm measurement. The red arrow represents the beam direction.}\label{fig:charm_detector_layout}
\end{figure*}

The detector layout of the SHiP-charm experiment was optimised in order to provide full topological and kinematic reconstruction of the event. A picture of the overall setup installed in the H4-PPE134 experimental area is shown in Figure~\ref{fig:charm_detector_layout}.

The topological reconstruction of proton interactions and the identification of charmed hadron decay vertices is performed within the target, which exploits the submicrometer and milliradian resolution of nuclear emulsions.
%In order to accomplish this challenging task, while maximising the 

\begin{figure}[htbp]
\centering
\includegraphics[width=1.0\linewidth]{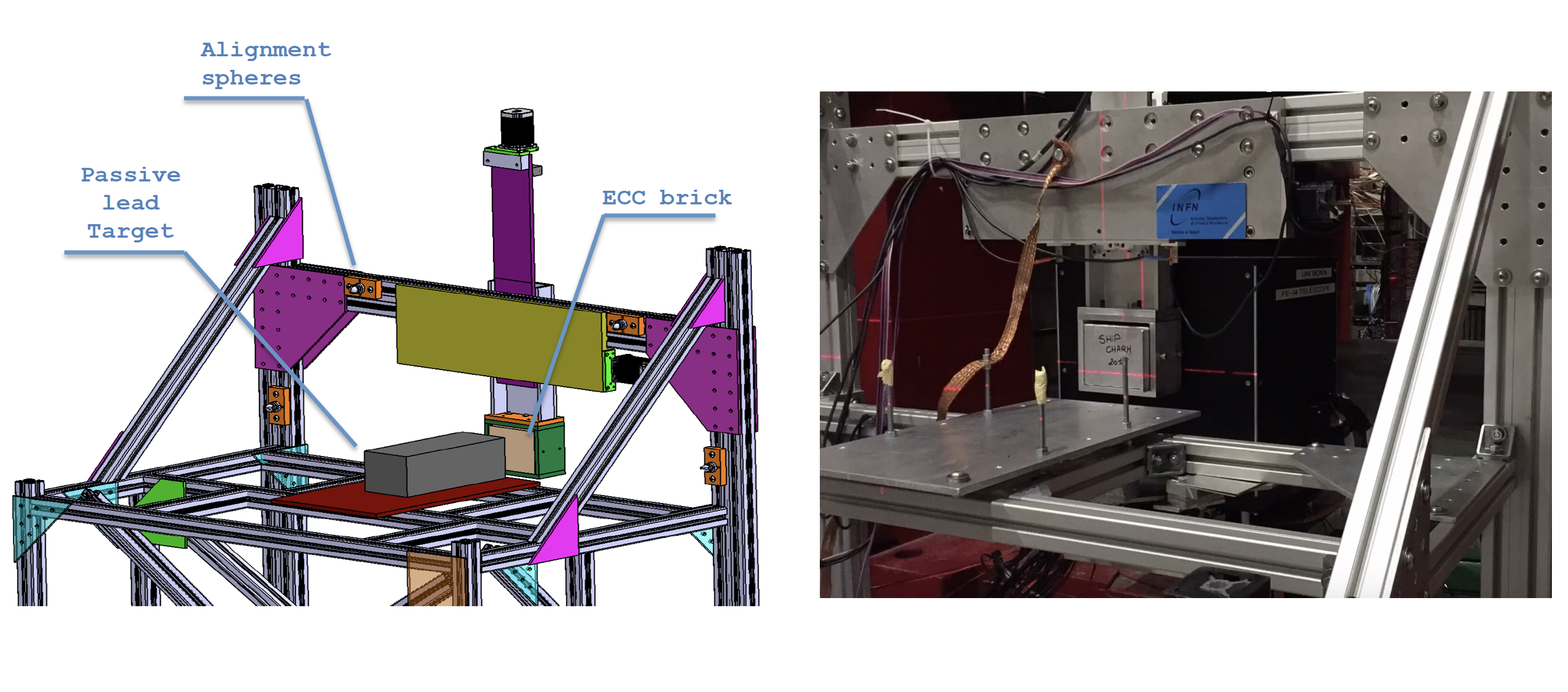}
\caption{Left: technical drawing of the target mover. Right: picture of the mechanical stage during a test exposure of an ECC target.}\label{fig:target_mover}
\end{figure}

The target is constructed according to the Emulsion Cloud Chamber (ECC) technique, alternating 1 mm-thick passive material plates with $330\,\upmu $m emulsion films. The ECC was placed on a motorised mechanical stage in order to ensure a uniform distribution of the proton beam over the whole emulsion surface of 125$\times$100~mm$^2$. A schematic drawing and a picture of the target mover are shown in Figure~\ref{fig:target_mover}. During each spill the target moves along the horizontal axis
($x$) at the uniform speed of 2.6 cm/s, thus covering the horizontal dimension of the ECC. Between two consecutive spills the target moves along the vertical axis ($y$) by 1 or 2 cm, depending on the expected track density in different target configurations. The
total target surface is consequently covered in 5 or 10 spills, respectively. 

A magnetic spectrometer is located downstream of the target. The magnetic field is provided by the GOLIATH magnet \cite{Rosenthal:2310483}, located in PPE134 area. In order to cope with the high multiplicity of tracks produced in each proton interaction, the upstream station is required to be highly segmented and withstand a high occupancy. Insertable B-Layer (IBL~\cite{Abbott:2018ikt}) hybrid silicon pixel detectors were used for this purpose. Pixels have a size of $250\times50$ $\upmu$m$^2$; pixel modules consist each of a planar sensor and two custom developed large FE-I4 front-end chips~\cite{GarciaSciveres:2011zz} with a sophisticated readout architecture. Each sensor is made of 160 columns and 336 rows, resulting in 53760 pixels.
The pixel tracking station is made of six planes equipped with IBL double-chips modules. Every second plane is rotated by 90$^\circ$ in order to provide a 50$\upmu$m position accuracy in both coordinates.
The upstream station covers a transverse area of about $33.6\times37.0$~mm$^2$, sufficient to contain the beam spot and proton interaction products passing through the lead-emulsion target. The matching between the emulsion target and the silicon pixel detector was successfully performed~\cite{SHiP:2021kxh}.

The downstream station is made by a combination of two different technologies: Scintillating fibers (SciFi) (T3s and T4s) in the central $40 \times 40$ cm$^2$ region, where the track density is higher, and drift tubes (T3 and T4) in the outer region.
The T3 and T4 stations consist each of four detection planes to provide XU and YV coordinates, where U and V planes have a
stereo angle of $\sim$2.5$^\circ$ with respect to X an Y, respectively. Each detector plane is made by $3 \times 12$~cm-wide mats of scintillating fibers~\cite{Beattie:2014ova}. A mat is a matrix structure consisting of six staggered fibre layers with a horizontal pitch of 270 $\upmu$m and a total length of 40 cm.

While the SciFi stations were built for the purpose of this measurement, drift tube chambers were adapted from  modules built for the OPERA experiment~\cite{Zimmermann:2006xr}. T3 and T4 stations provide the $x$-coordinate information in the external region downstream of the GOLIATH magnet. Drift tube modules were installed on both  sides and above the region covered by the SciFi stations.

The most downstream component of the experiment is the Muon Filter, which is designed to identify muons with high efficiency, separating them from charged hadrons.
At the same time, it has to reconstruct the muon track slope to match the corresponding track reconstructed in the upstream Magnetic Spectrometer and assign the momentum to the muon track. 
The Muon Filter consists of five concrete slabs, two 80 cm-thick and three 40 cm-thick, acting as hadron absorbers,  interleaved with five Resistive Plate Chambers (RPCs), acting as trackers. The transverse size of the RPC planes is $195 \times 125$~cm$^2$. The muon identification is performed on the basis of the number of crossed layers in the detector. The RPCs  were designed and constructed to operate in avalanche mode, with a time resolution of about 1 ns. Two orthogonal sets of strips,  1 cm-wide, are used for 2D measurements with a position resolution of about 3~mm in both directions.

%% file: sections/exposure.tex
%\section{Data taking}
\section{Data taking and simulation}

\begin{figure*}
\centering
\includegraphics[width=0.9\linewidth]{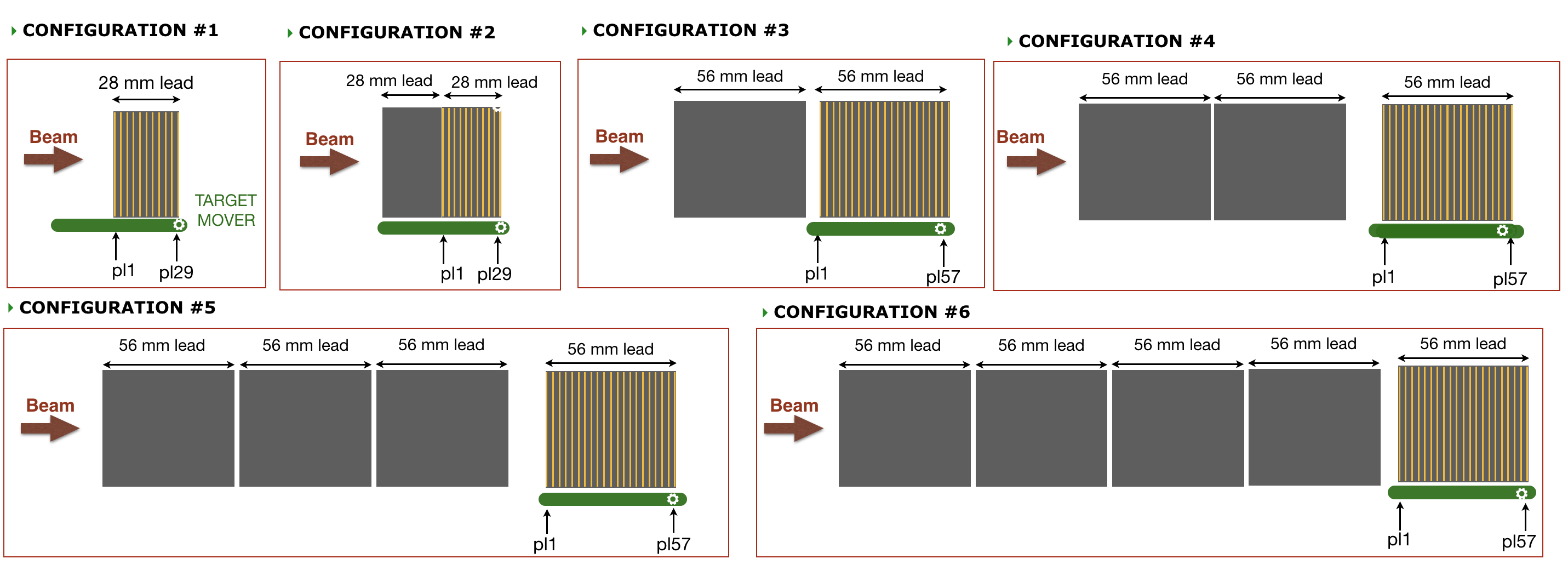}
\caption{Schematic layout of the six target configurations.}\label{fig:charm_configuration}
\end{figure*}%

The SHiP-charm optimisation run was performed in July 2018. The target was assembled in six different configurations in order to study the production of charmed hadrons at different depths, up to a total thickness of 280 mm, corresponding to about 1.6 interaction lengths.
The most downstream section of the target is instrumented with nuclear emulsions (the ECC) and moved by the motorised stage.

Upstream of the ECC, lead blocks with lengths from 28 to 244 mm are positioned to act as a pre-shower, according to the scheme shown in Figure~\ref{fig:charm_configuration}. Hereafter the six target configurations will be referred to as CHARM$n$, with $n$ ranging from 1 to 6.

The ECC target of CHARM1 and CHARM2 is made of a sequence of 29 emulsion films alternated with 28 passive layers, while for configurations from CHARM3 to CHARM6 it consists of 57 emulsion films and 56 passive layers. Multiple runs were performed for the different configurations (CHARM$x$-RUN$m$, with $m$ ranging from 1 to 6) in order to accumulate enough statistics in each portion of the target. A total number of $15.6 \times 10^5$ p.o.t. were integrated during the whole exposure. 
All runs used lead as passive material, except for the sixth run of CHARM1, which used 1 mm-thick tungsten layers. 
%The composition of each configuration, the number of runs and the number of integrated p.o.t. are summarised in Tab.~\ref{tab:exposure}.

\begin{comment}
\begin{table*}[htbp]
\centering
\begin{tabular}{c|c|c|c|c|c}
\toprule
Configuration & n Runs  & Pre-shower & ECC & n Films & integrated  \\
& &   &  &&  p.o.t. [$10^5$]\\
\midrule
CHARM 1 & 6 & / & 28 mm Pb(W) + 29  films & 174 & 5.4 \\
CHARM 2 & 6 & 28 mm Pb & 28 mm Pb + 29  films & 174  & 5.2\\
CHARM 3 & 3 & 56 mm Pb & 56 mm Pb + 57  films & 171 & 1.0 \\
CHARM 4 & 3 & 113 mm Pb & 56 mm Pb + 57  films & 171 & 0.8\\
CHARM 5 & 3 & 168 mm Pb & 56 mm Pb + 57  films & 171 & 1.6\\
CHARM 6 & 3 & 224 mm Pb & 56 mm Pb + 57  films & 171 & 1.6 \\
\midrule
TOTAL & 24 & & & 1032 & 15.6\\
\bottomrule
\end{tabular}
\caption{Summary of the SHiP-charm 2018 exposure.}
\label{tab:exposure}
\end{table*}
\end{comment}

A total amount of 1032 emulsion films were used, corresponding to $\sim$12 m$^2$. They were produced by Nagoya University and Slavich Company in June 2018. 
Emulsion films consist of two 75~$\upmu$m-thick layers of nuclear emulsion, separated by a 180~$\upmu$m-thick plastic base.
The transverse size is $125 \times100$ mm$^2$. ECC targets were assembled in a dedicated facility at CERN right before the exposure. %The exposure was performed at room temperature. 
After the exposure, targets were transferred to the CERN facility, disassembled,  and  emulsion films underwent chemical treatment. 

\begin{figure}
\centering
\includegraphics[width=1.0\linewidth]{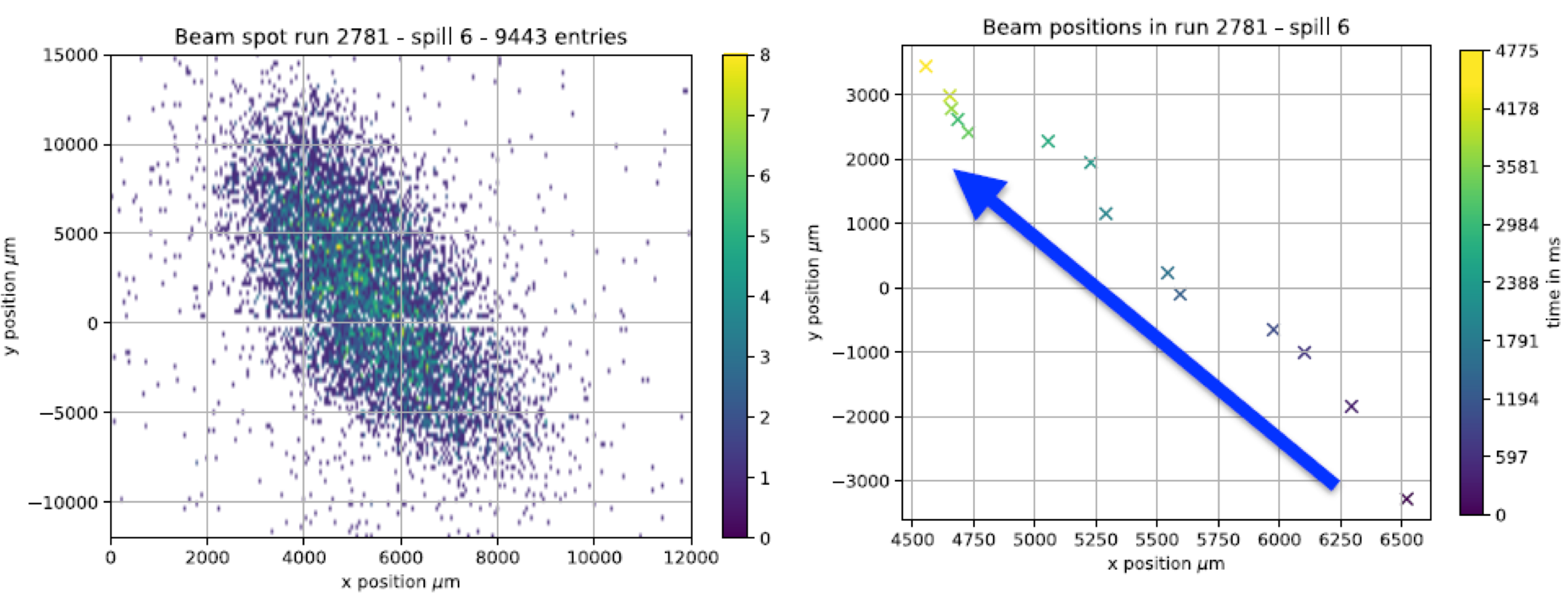}
\caption{Left: beam profile in the transverse plane, as registered by the pixel detector in the sixth spill of CHARM2-RUN1. Right: position of the beam center-of-gravity as function of time during  the spill.}\label{fig:profile}
\end{figure}

The proton beam intensity was measured by a beam counter located upstream of the target region. The temporal structure of the beam was consistent during the whole exposure, with a spill duration of 4.8 s. Its intensity, however, showed fluctuations from $7.7 \times10^3$ to $13.8\times10^3$ protons/spill. The profile of the beam during the spill was monitored by the pixel station. The beam profile recorded in one spill is shown in Figure \ref{fig:profile}. The beam spot integrated during the spill has a transverse size of about $6\times15$ mm$^2$. The elliptical shape is due to a translation of the beam center-of-gravity within the spill. 

%% file: sections/simulation.tex
%\section{Simulation}
%\label{sec:simulation}

\begin{figure}[htbp]
\centering
\includegraphics[width=1.0\linewidth]{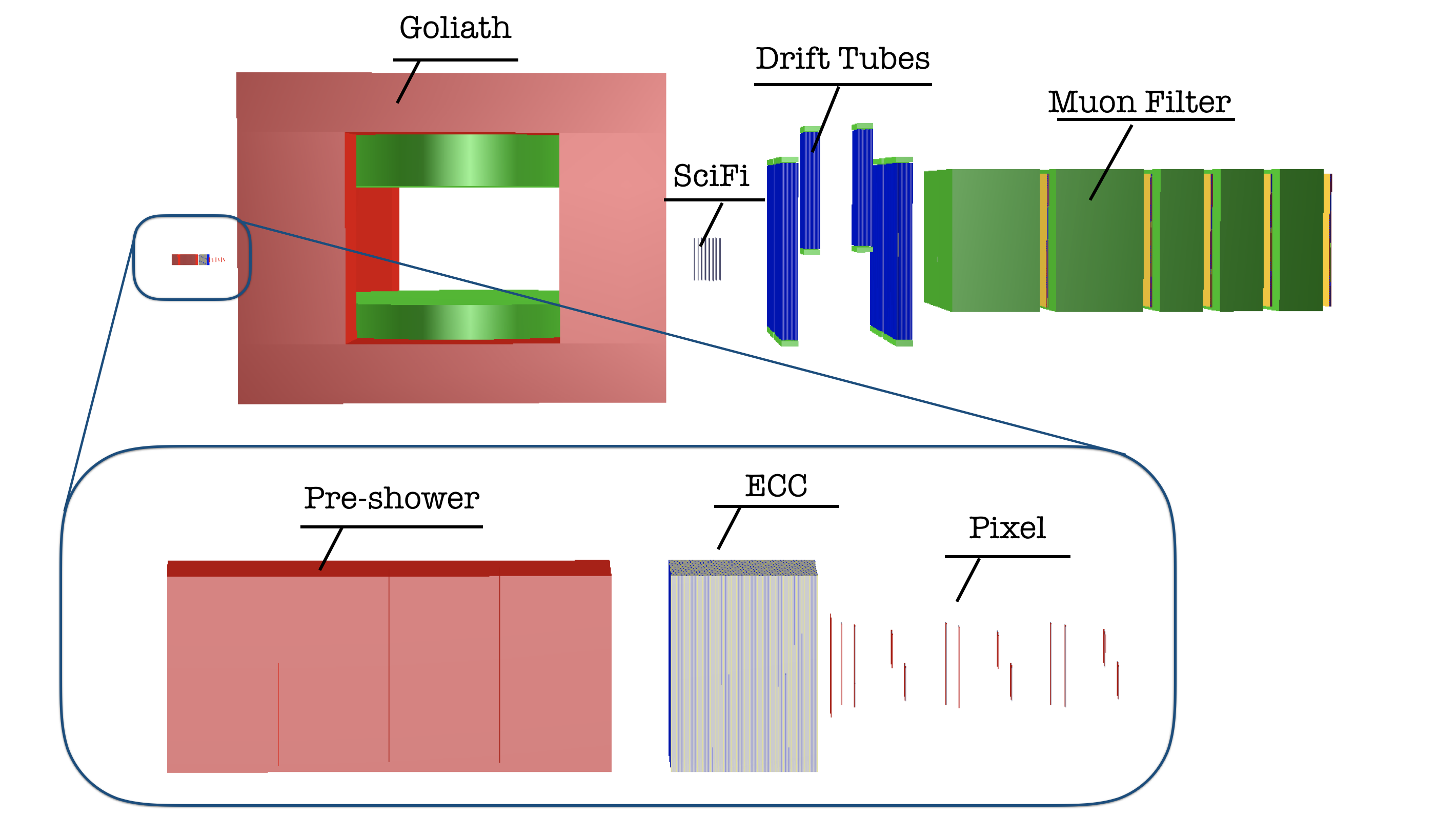}
\caption{Layout of the SHiP-charm experimental layout, as implemented in FairShip.}\label{fig:charm_simulation}
\end{figure}%

The SHiP-charm experimental apparatus was reproduced within the FairShip software, the official SHiP simulation framework derived from FairRoot \cite{fairroot}, as shown in Figure~\ref{fig:charm_simulation}. The geometry and the position of different sub-detectors were set taking into account measurements performed in situ by the CERN survey team. 
The magnetic-field map measured by the CERN staff in 2017 \cite{Rosenthal:2310483} was imported in the simulation of the GOLIATH magnet.
The simulation of 400 GeV/c proton interactions within the target and the propagation of particles in detector materials is performed with GEANT4 \cite{Agostinelli:2002hh, Allison:2016lfl}, that was validated to provide reliable estimates in the SHiP-charm energy regime~\cite{Ivanchenko:2017rvi,SHiP:2020hyy}.
Different simulation campaigns were performed in order to reproduce the six target configurations.

%% file: sections/event_reconstruction.tex
\section{Data analysis}
\label{sec:reconstruction}

\subsection{Track reconstruction in nuclear emulsions}
\begin{figure}[htbp]
\centering
\includegraphics[width=1.0\linewidth]{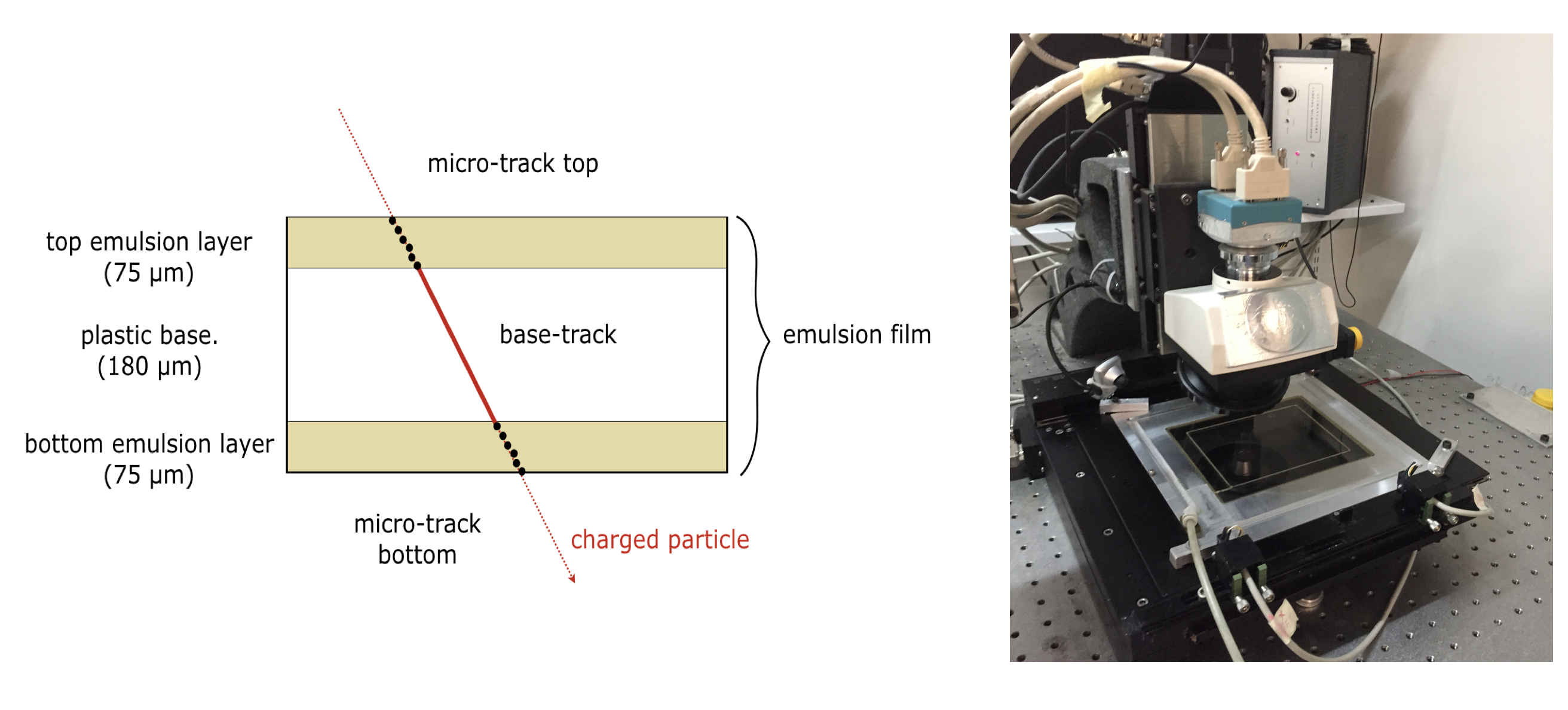}
\caption{Left: schematic layout of a nuclear emulsion film. Right: one of the optical microscopes used for the analysis of nuclear emulsions exposed in the SHiP-charm project.}\label{fig:emulsion}
\end{figure}

The track left by a charged particle in an emulsion layer is recorded by a series of sensitised AgBr crystals, growing up to 0.6 $\upmu$m diameter during the development process. Optical microscopes analyse the whole thickness of the emulsion, acquiring tomographic images at equally spaced depths. The acquired images are digitized, then an image processor recognizes the grains as  \textit{clusters}, i.e.~groups of pixels of given size and shape. Then, a track in the emulsion layer (usually referred to as \textit{micro-track}) is obtained connecting clusters belonging to different levels, as shown in the left panel of Figure~\ref{fig:emulsion}. Since an emulsion film is formed by two emulsion layers, the connection of the two micro-tracks through the plastic base provides a reconstruction of the particle's trajectory in the emulsion film, called \textit{base-track}.  
The reconstruction of particle tracks in the full volume requires connecting base-tracks in consecutive films. In order to define a global reference system, a set of affine transformations has to be computed to account for the different reference frames used for data taken in different films.

Once all emulsion films are aligned, \textit{volume-tracks} (i.e., charged tracks which crossed several emulsion films) can be reconstructed. The track finding and fitting is
 based on the Kalman-Filter algorithm and takes into account possible inefficiencies in the base-track reconstruction \cite{Arrabito:2007td}.

The vertex identification is initiated by two-track vertices defined according to minimal distance criteria. Topological cuts are used in order to reduce the combinatorial background. The final selection on the track pairs is based on a vertex probability calculated with the full covariance matrix of the involved tracks. Starting from pairs, $n$-tracks vertices are constructed using the Kalman-Filter technique.
The off-line reconstruction tool used in the analysis reported in this document is \mbox{FEDRA} (Frame-work for Emulsion Data Reconstruction and Analysis) \cite{Tyukov:2006ny}, an object-oriented tool based on C++ language and developed in the ROOT \cite{Brun:2000es} framework.

The analysis of emulsion films was performed in dedicated laboratories in Naples and Zurich
equipped with a new generation of optical microscopes, one of
which is shown in the right panel of Figure~\ref{fig:emulsion}. A recently developed upgrade of the European Scanning System (ESS) \cite{Armenise:2005yh, Arrabito:2006rv, DeSerio:2005yd} was used.  The use of a faster camera with smaller sensor pixels and a higher number of pixels combined with a lower magnification objective lens, together with a new software LASSO \cite{Alexandrov:2016tyi,Alexandrov:2015kzs} has allowed to increase the scanning speed to 180 cm$^2$/h \cite{Alexandrov:2017qpw}, more than a factor ten larger than the previous generation.

\subsection{Proton-beam characterisation}

The number of protons impinging on ECC target units vary from 10$^2$/cm$^2$ to 10$^3$/cm$^2$ according to the configuration of the exposure. The data analysis shows that the track density increases with the depth in the brick due to the proton interactions, hadronic reinteractions and electromagnetic showers.
%, as shown in Figure~\ref{fig:charm_density}. 
The density of segments reconstructed in a single emulsion film extends up to 4.5$\times$10$^4$/cm$^2$.  
%The challenge of the SHiP-charm measurement is two-fold: reconstruct tracks and interaction vertices in a high density environment, and search for rare decays of charmed hadrons. 

\begin{figure}
\centering
\includegraphics[width=1.0\linewidth]{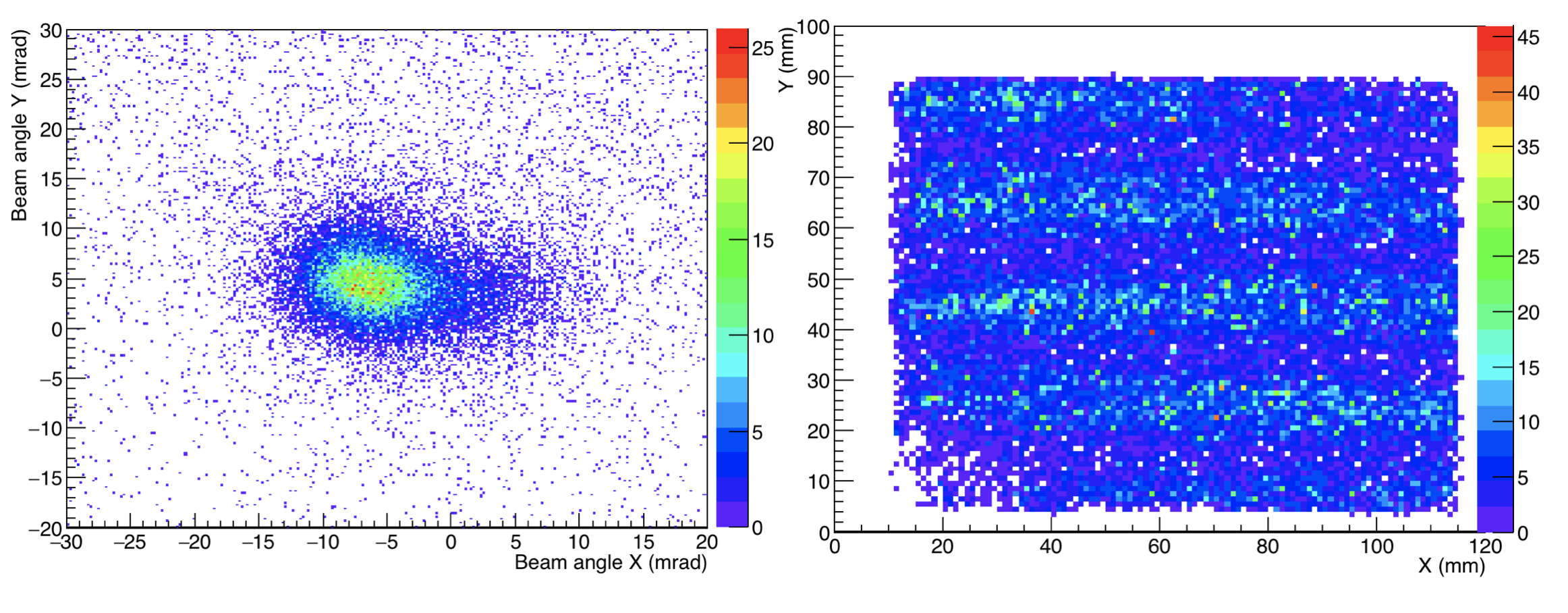}
\caption{Left: angular dispersion of the proton beam as reconstructed in one of the exposed ECC target units. Right: position distribution of incoming protons on the emulsion surface.}
\label{fig:charm_beam}
\end{figure}

\begin{figure}
\centering
\includegraphics[width=0.9\linewidth]{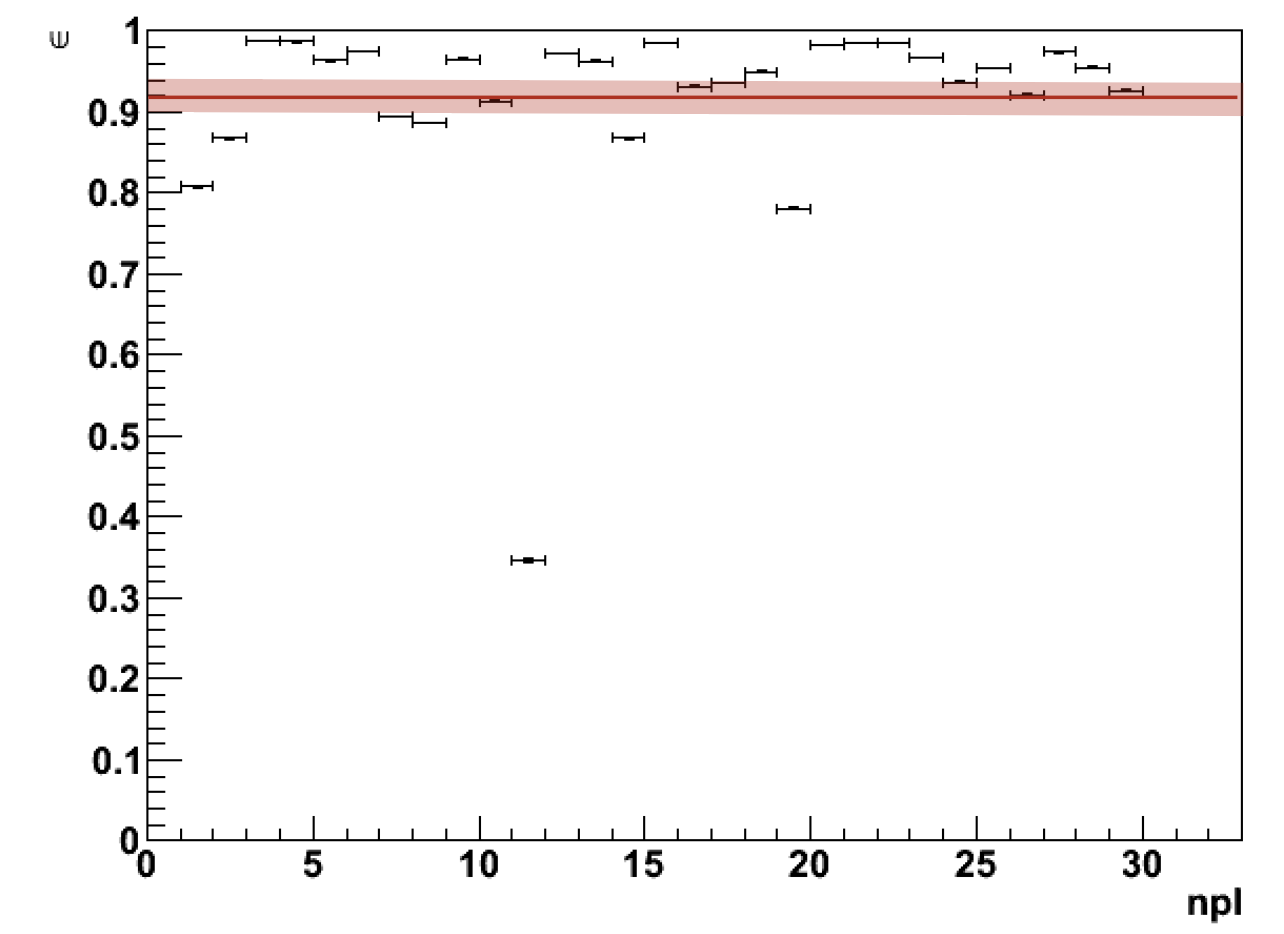}
\caption{Film-by-film base-track efficiency as a function of the film number for reconstructed protons in CHARM1-RUN2 configuration. The average efficiency, amounting to $92\pm2$ \%, is shown as horizontal red line. }
\label{fig:track_eff}
\end{figure}

%Each ECC was mounted on a moving table in order to change the position of the target with respect to the proton beam and irradiate the whole surface of the detector. 
Figure ~\ref{fig:charm_beam} shows the characterisation of the proton beam in one of the ECC targets both in terms of angle (left) and position (right). The pattern observed in the position distribution reproduces the movement of the target with respect to the proton beam.
The base-track efficiency is shown in Figure~\ref{fig:track_eff} as a function of the film number in one of the most upstream configurations. The average base-track efficiency is higher than 90\%. A slight decrease in the efficiency is observed in downstream configurations due to higher track density.

\begin{comment}
\begin{figure}
\centering
\includegraphics[width=1.0\linewidth]{figs/charm_density1.png}
\caption{Left: tracks reconstructed in a 1$\times$1 cm$^2$ of the configuration CHARM1-RUN6. Right: track density in one of the most downstream target units.}
\label{fig:charm_density}
\end{figure}%
\end{comment}

%% file: sections/vertex_identification.tex
\section{Interaction-vertex identification}

Several thousands of proton interaction vertices are expected in a single target unit ($\sim$10$^3$ cm$^3$). 400 GeV/c proton interactions produce on average more than ten charged particles and as many photons, having energies up to $\sim$50 GeV. This results in a large number of secondary hadronic re-interactions and electromagnetic showers, that increases the number of reconstructed vertices by two order of magnitudes.
To set the scale,  the unitary cell of the OPERA experiment \cite{Acquafredda:2009zz,Agafonova:2018dkb} contained only a single neutrino-interaction vertex in the same volume.

The analysis of the SHiP-charm emulsion data therefore required the development of dedicated software and analysis tools to extract the signal from an unprecedented background rate. The main background source consists of low energy particles produced in hadronic and electromagnetic showers originated in primary protons interaction and in subsequent re-interactions downstream of the primary vertex. The yield of background vertices in the reconstructed sample is more than one order larger than the signal and shows an increasing trend in downstream configurations.

A full Monte Carlo simulation was performed in order to have a training sample that accurately reproduced data. The tracking and vertexing algorithms described in Section \ref{sec:reconstruction} were applied both on simulated and real data. 

\begin{comment}
Distributions shown in Figure~\ref{fig:rec_vtx} show that the simulation reproduces the data fairly well for multiplicities larger than six.

\begin{figure}
\centering
\includegraphics[width=1.0\linewidth]{figs/vz_mult_no_bdt}
\caption{Charged track multiplicity (left) and position distribution along the beam axis (right) for vertices reconstructed in CHARM1-RUN1 configuration. Data points are shown in red, simulation is represented in blue. Distributions have been normalised to the number of p.o.t. integrated in the analised run.}
\label{fig:rec_vtx}
\end{figure}%

\end{comment}

A multivariate classification was performed using boosted
decision trees from the TMVA toolkit \cite{Voss:2007jxm} to distinguish a signal of true interaction vertices from the background. The
background is mainly due to the random combination of low-momentum tracks and electromagnetic showers that crowd the ECC volume. 
Five discriminating variables were selected:
\begin{itemize}
    \item vertex probability, as provided by the fit procedure 
    \item angular distance between tracks associated to the vertex
    \item mean impact parameter of tracks at the vertex
    \item maximum impact parameter of tracks at the vertex
    \item fill factor of tracks at the vertex, defined as the ratio between the number of base-tracks building up the track and the number of emulsion films downstream of the vertex.
\end{itemize}

Other variables were tested, such as the average slope of tracks associated to the vertex and distribution of tracks in the transverse plane, and resulted not to have good performances in the signal-to-background discrimination.

\begin{figure}
\centering
\includegraphics[width=1.0\linewidth]{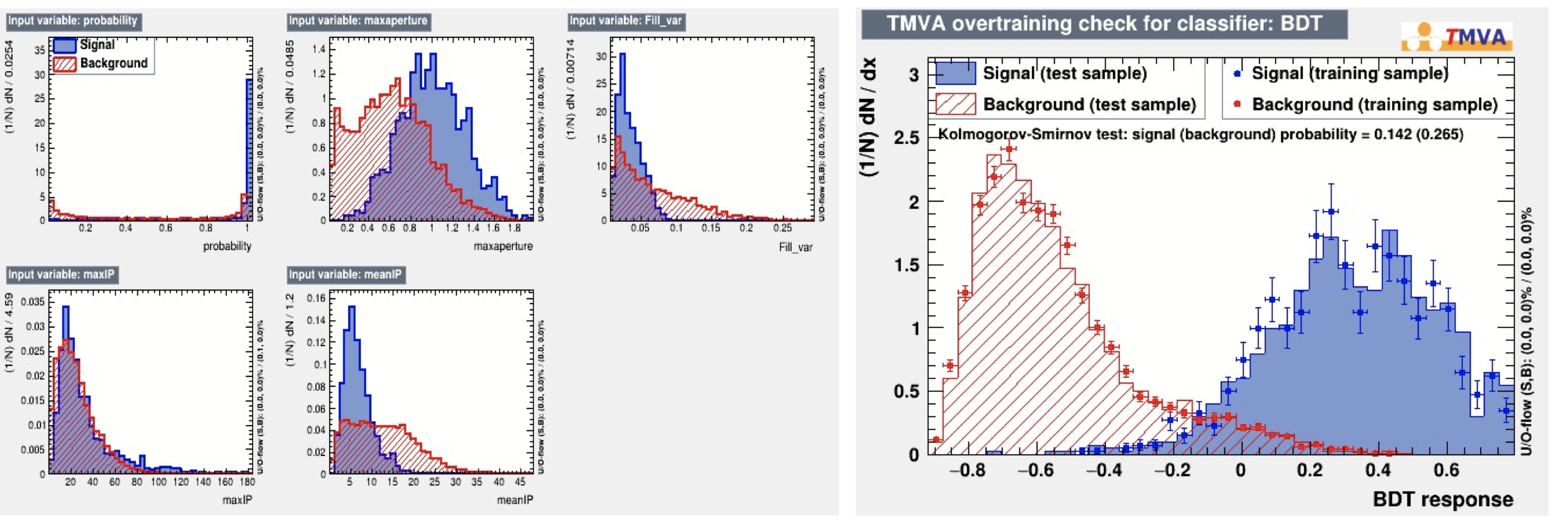}
\caption{Left: distribution of input variables used in the multivariate analysis. Right: output value of the BDT for signal (blue) and background (red).}
\label{fig:bdt}
\end{figure}

\begin{figure}[htbp]
\centering
\includegraphics[width=1.0\linewidth]{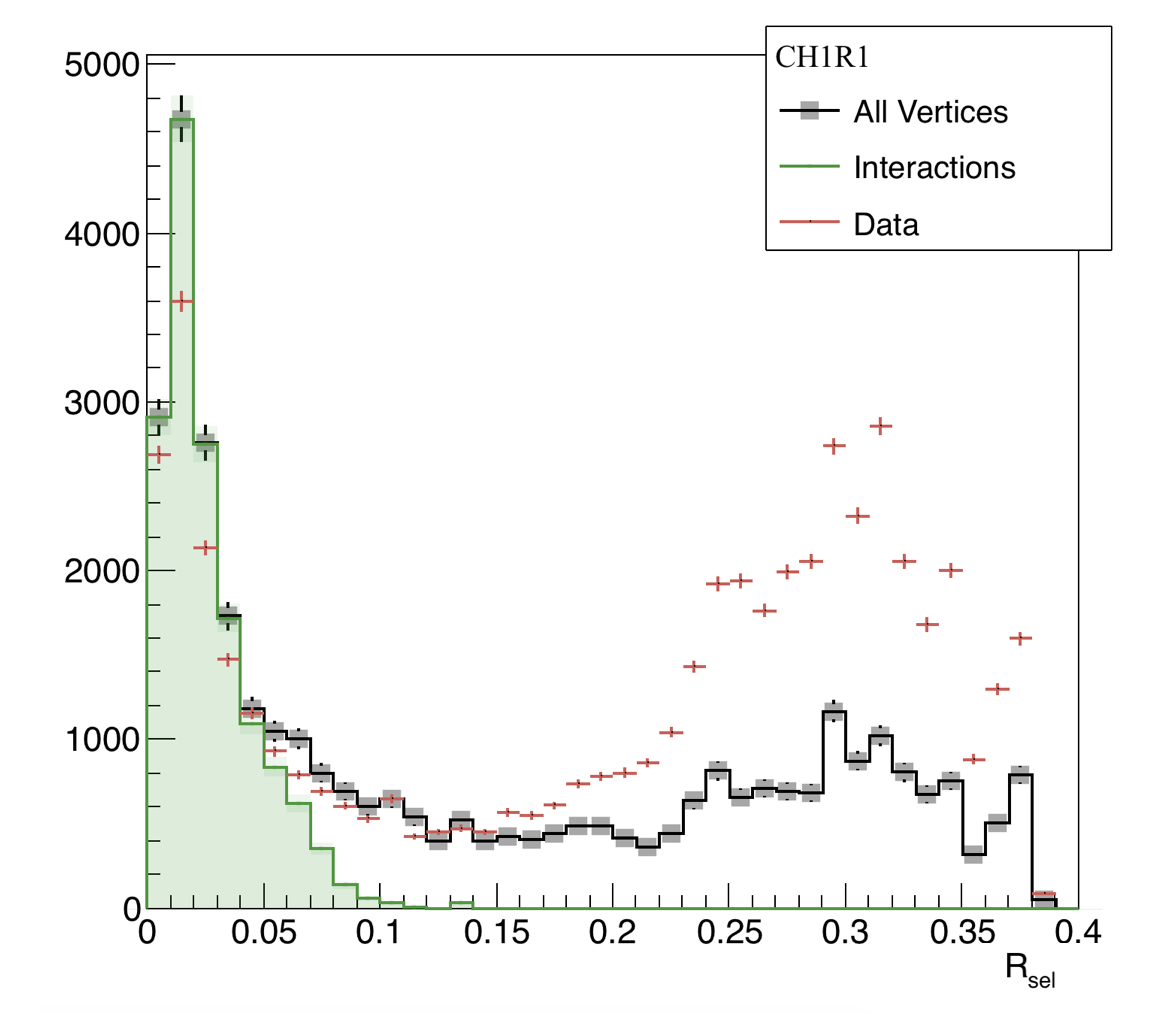}
\caption{Distribution of the $R_{\mathrm{sel}}$ variable for data and simulated signal and background vertices.}
\label{fig:r_sel}
\end{figure}

The left panel of Figure~\ref{fig:bdt} shows the above mentioned variables for the training sample. The output of the BDT ($V_{\mathrm{bdt}}$) is shown in the right panel of Figure~\ref{fig:bdt}: a good separation between signal and background distributions is observed. The final selection of the signal component is performed on the variable $R_{\mathrm{sel}}$, defined as the ratio between  (1-$V_{\mathrm{bdt}}$) and the track multiplicity at the reconstructed vertex. The distribution of $R_{\mathrm{sel}}$ variable is shown in  Figure~\ref{fig:r_sel} for data and simulation. The signal component is confined in the region $R_{\mathrm{sel}}<0.1$, where a fairly good agreement between data and simulation is observed. The excess in the data for higher $R_{\mathrm{sel}}$ values is due to very low ($n<4$) multiplicity vertices that are mainly made of random combination of instrumental background tracks. This background component, indeed, is not included in the current version of the simulation software.
The cut on the $R_{\mathrm{sel}}$ variable was optimised in order to maximise the background rejection while keeping an high signal selection efficiency. Vertices having $R_{\mathrm{sel}}<0.05$ are  classified as interaction-vertex candidates.

\begin{figure}[htbp]
\centering
\includegraphics[width=0.8\linewidth]{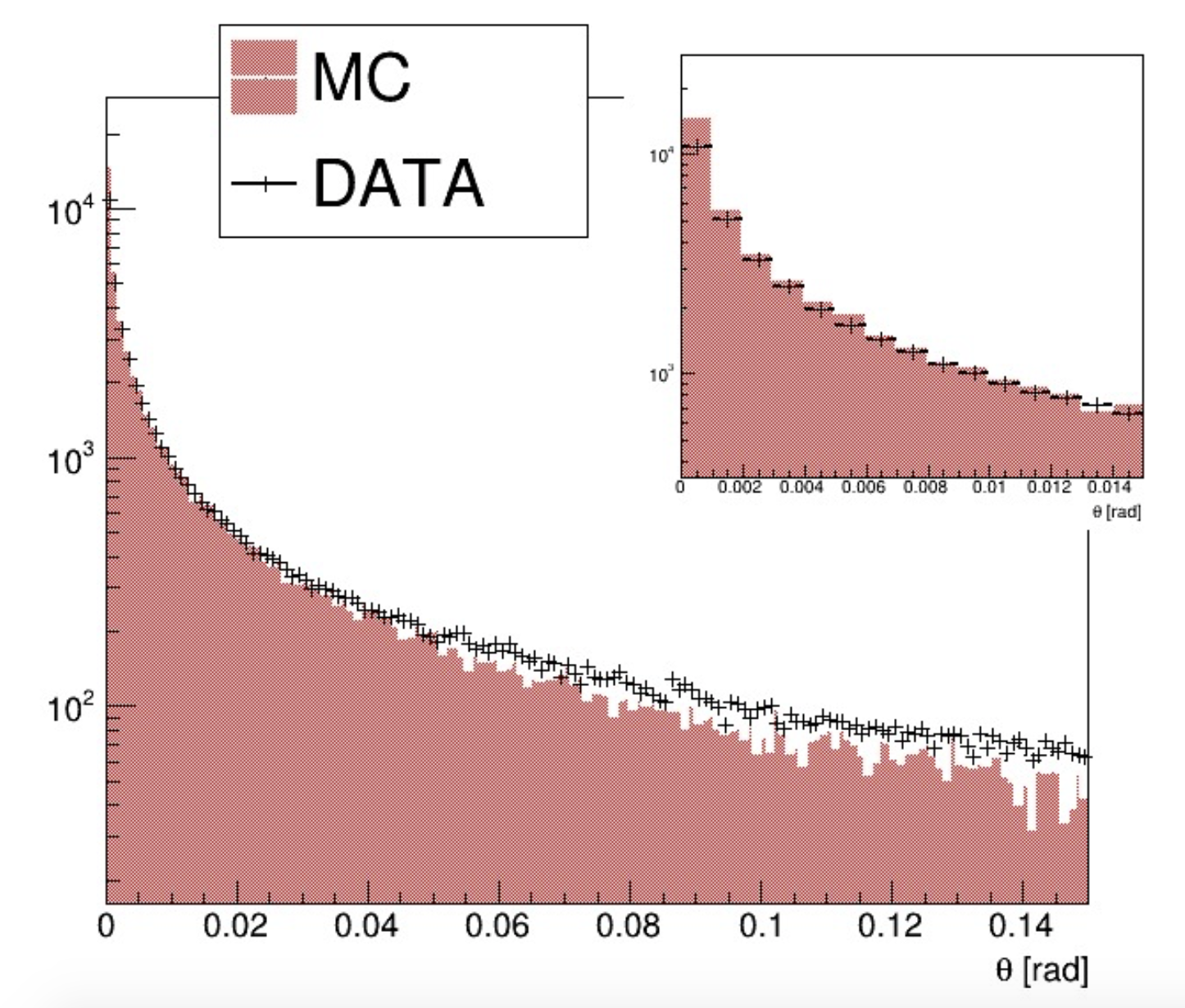}
\caption{Angular distribution of tracks associated to interaction-vertex candidates. The inset shows the region with slopes smaller than 0.014 rad.}
\label{fig:slope}
\end{figure}

The angular distribution of tracks associated to interaction-vertex candidates is shown in Figure~\ref{fig:slope}. A good agreement is observed, both in normalisation and shape, thus validating the Monte Carlo simulation and the signal selection procedure.

\section{Interaction-vertex characterisation}

\begin{figure}
\centering
\includegraphics[width=1.0\linewidth]{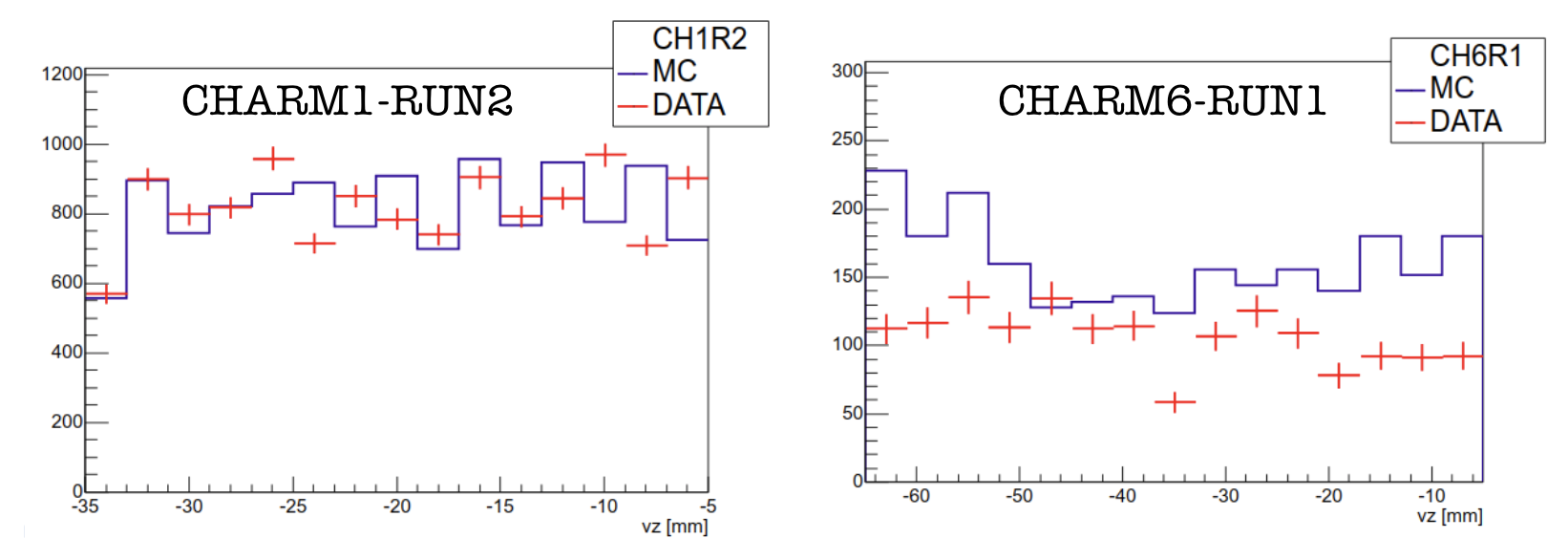}
\caption{Vertex position along the beam direction for interaction-vertex candidates reconstructed in CHARM1-RUN2 (left) and CHARM6-RUN1 (right). Data and Monte Carlo distributions have been normalised to the number of p.o.t. integrated in the analised run.}
\label{fig:charm_primaryZ_sel}
\end{figure}

The reconstructed position of interaction-vertex candidates along the beam direction for the most upstream and the most downstream configuration is shown in Figure~\ref{fig:charm_primaryZ_sel}. The most upstream configuration shows very good agreement between data and Monte Carlo, both in normalisation and shape. A discrepancy between data is observed in downstream configurations which is due to inefficiencies in track reconstruction that affect the overall number of selected vertices without introducing relevant biases in the variables that characterise interaction-vertex candidates. The origin of the discrepancies are mainly related to an higher track density in  downstream configurations, mainly coming from the overlap of hadronic and electromagnetic showers started in upstream regions. The presence of a large number of low-energy tracks can spoil the alignment precision between consecutive films, thus causing a discrepancy with respect to simulations, where those effects are not taken into account.

\begin{figure}
\centering
\includegraphics[width=1.0\linewidth]{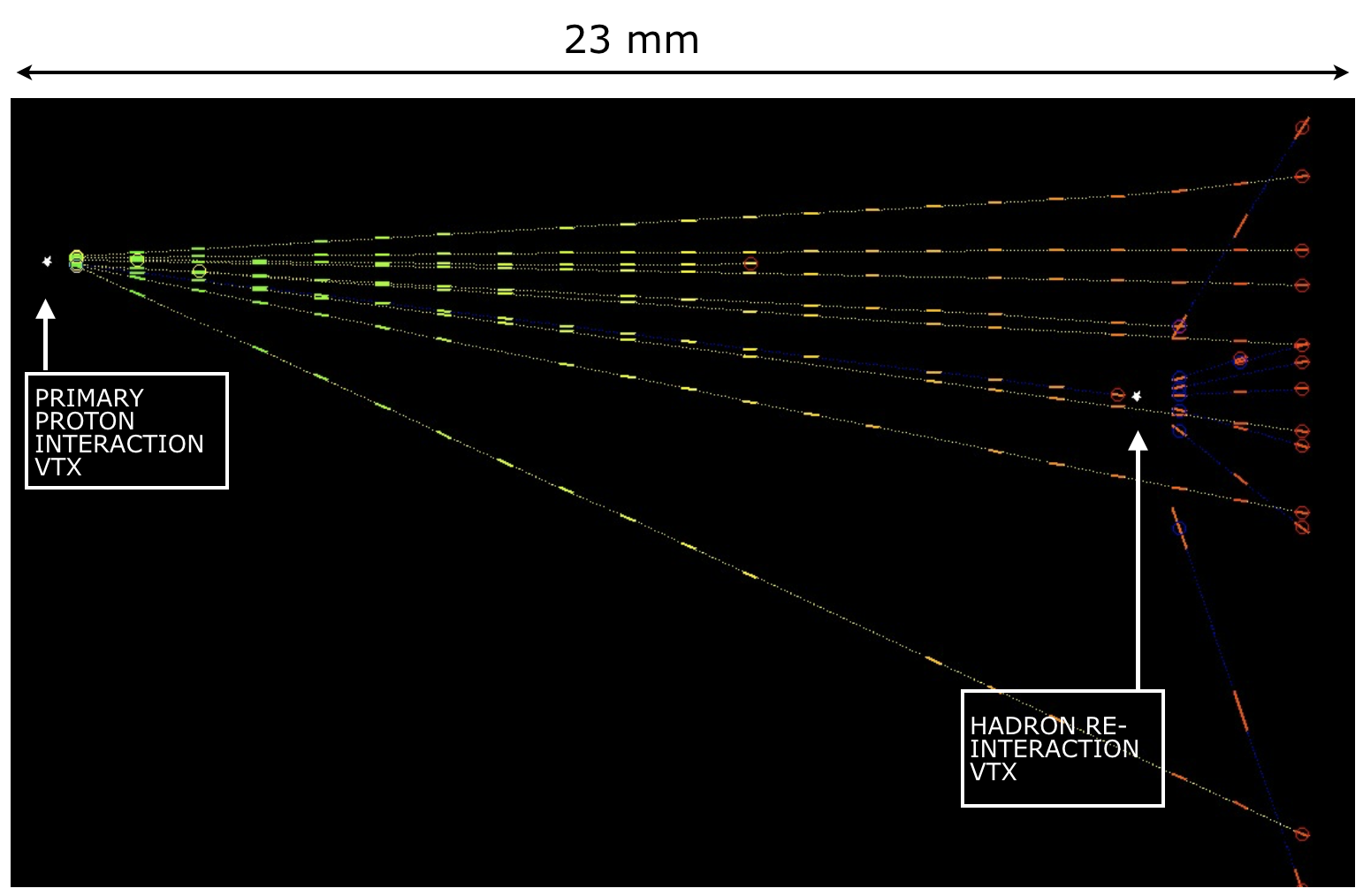}
\caption{Display of a reconstructed Monte Carlo event where both the primary-proton interaction vertex and an hadron-reinteraction vertex are reconstructed.}
\label{fig:signal_vertex}
\end{figure}

The signal sample selected with the above mentioned procedure is made of two components: primary protons interaction vertices and hadron re-interaction vertices. A display of a Monte Carlo event containing both vertex categories is shown in Figure~\ref{fig:signal_vertex}.

\begin{figure}
\centering
\includegraphics[width=1.0\linewidth]{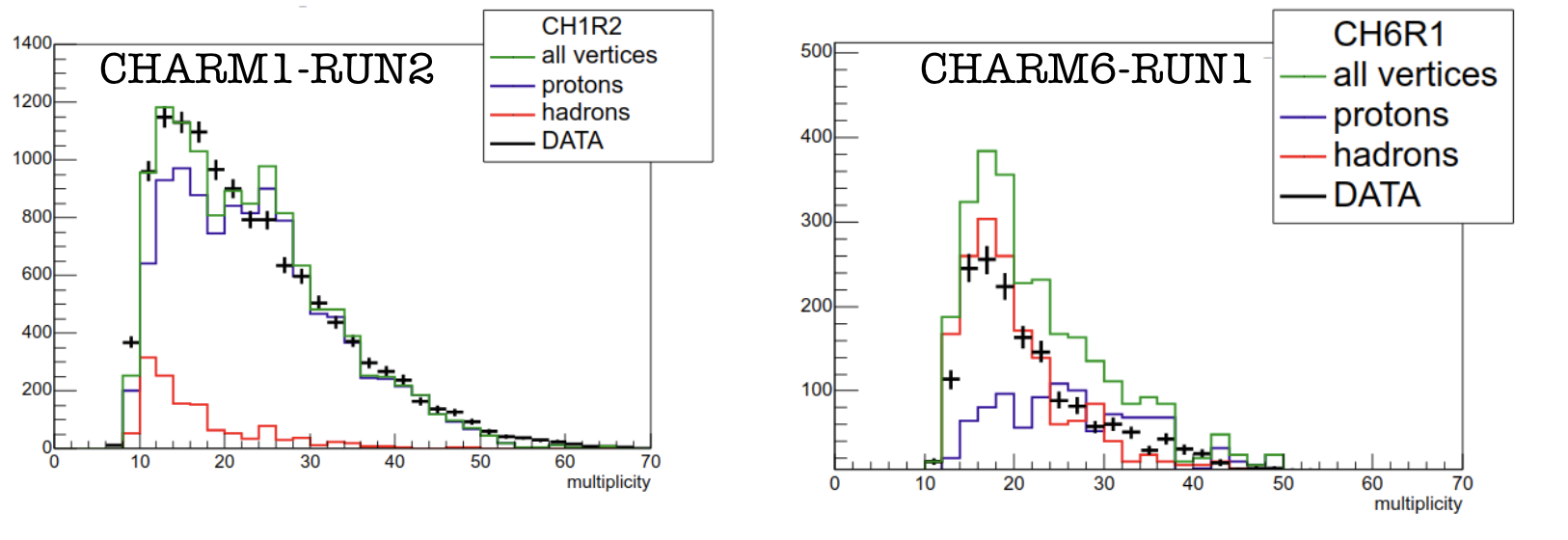}
\caption{Charged track multiplicity for interaction-vertex candidates reconstructed in CHARM1-RUN2 (left) and CHARM6-RUN1 (right). Data and Monte Carlo distributions have been normalised to the number of p.o.t. integrated in the analysed run.}
\label{fig:charm_primaryvertex_sel}
\end{figure}

The interaction vertex multiplicity for the most upstream and the most downstream configuration is shown in Figure~\ref{fig:charm_primaryvertex_sel}. The contribution of the primary proton and hadron-reinteraction components is shown separately. As one might expect, the hadron-reinteraction component increases as the configuration number increases, going from 11\% in CHARM1 to 59\% in CHARM6.

%one of the ECC target units are reported in Fig.~\ref{fig:charm_primaryvertex}. 

%A good agreement between data and Monte Carlo expectations is found, both in normalisation and in shape, for the number of charged tracks defining the interaction vertex and the position of the vertex along the beam axis. These results prove the capability to reconstruct interaction vertices in a harsh environment. 

%\begin{figure}[htbp]
%\centering
%\includegraphics[scale=0.20]{figs/vertex}
%\caption{Proton interaction vertices reconstructed in the ECC brick.}\label{fig:vertex}
%\end{figure}%

\begin{comment}
The list of configurations used for the analysis described in this document is reported in Tab.~\ref{tab:eff} together with measured efficiencies. The observed fluctuations are related to  different emulsion batches, handling procedures and chemical treatments used for the different runs.

\begin{table}[h]
\centering
 \begin{tabular}{cc | cc} 
 \toprule
 Configuration & Efficiency (\%) \quad & \quad Configuration & Efficiency (\%) \\ 
 \midrule
CHARM1-RUN1  &  83 \quad& \quad CHARM2-RUN4  &  55 \\
CHARM1-RUN2  &  99 \quad& \quad CHARM3-RUN1  &  70 \\
CHARM1-RUN4  &  53 \quad& \quad CHARM4-RUN1  &  38 \\
CHARM1-RUN5  &  49 \quad& \quad CHARM5-RUN1  &  51 \\
CHARM2-RUN2  &  57 \quad& \quad CHARM6-RUN1  &  66 \\
CHARM2-RUN3  &  41 \quad& \quad & \\
 \bottomrule
 \end{tabular}
 \caption{Vertex reconstruction efficiencies measured in the analysed configurations.}  \label{tab:eff}
\end{table}

\end{comment}

%% file: sections/results.tex
\section{Results}
In order to merge data reconstructed in different configurations, inefficiencies  were corrected by applying a normalisation factor, which also scaled all data to the same number of incoming protons on target.

\begin{comment}
For a run CHARM$x$-RUN$y$ with a configuration index $x$ and a run index $y$, the normalisation factor $f_{xy}$ 
%$1/\epsilon_{xy}$ 
can be evaluated as:

\begin{equation}
	%\frac{1}{\epsilon_{xy}}=
	f_{xy}= 
	c_x \sum_y \frac{N_{xy} \cdot \epsilon_{xy}}{N_{tot}}
\end{equation}

where $N_{xy}$ is the number of p.o.t., $N_{tot}$ is ??? and $\epsilon_{xy}$ is efficiency measured in each run. $c_x$ is the same for all runs of the same target configuration $x$. It is used to normalise all configurations to the same number of protons on target, using the simulation of CHARM1 as as reference, and can be defined as 

\begin{equation}
	c_x = \frac{N^{MC1}_{pot}}{N_{pot}^{MCx}} \cdot \frac{1}{Nx}
\end{equation}

where $N_{pot}^{MCx}$ is the number of $p.o.t.$ simulated for the configuration $CHx$ and ${Nx}$ is the number of analised runs for that specific configuration.

\end{comment}

%using the Monte Carlo from CHARM1 as reference:

\begin{figure*}[htbp]
\centering
\includegraphics[width=0.7\linewidth]{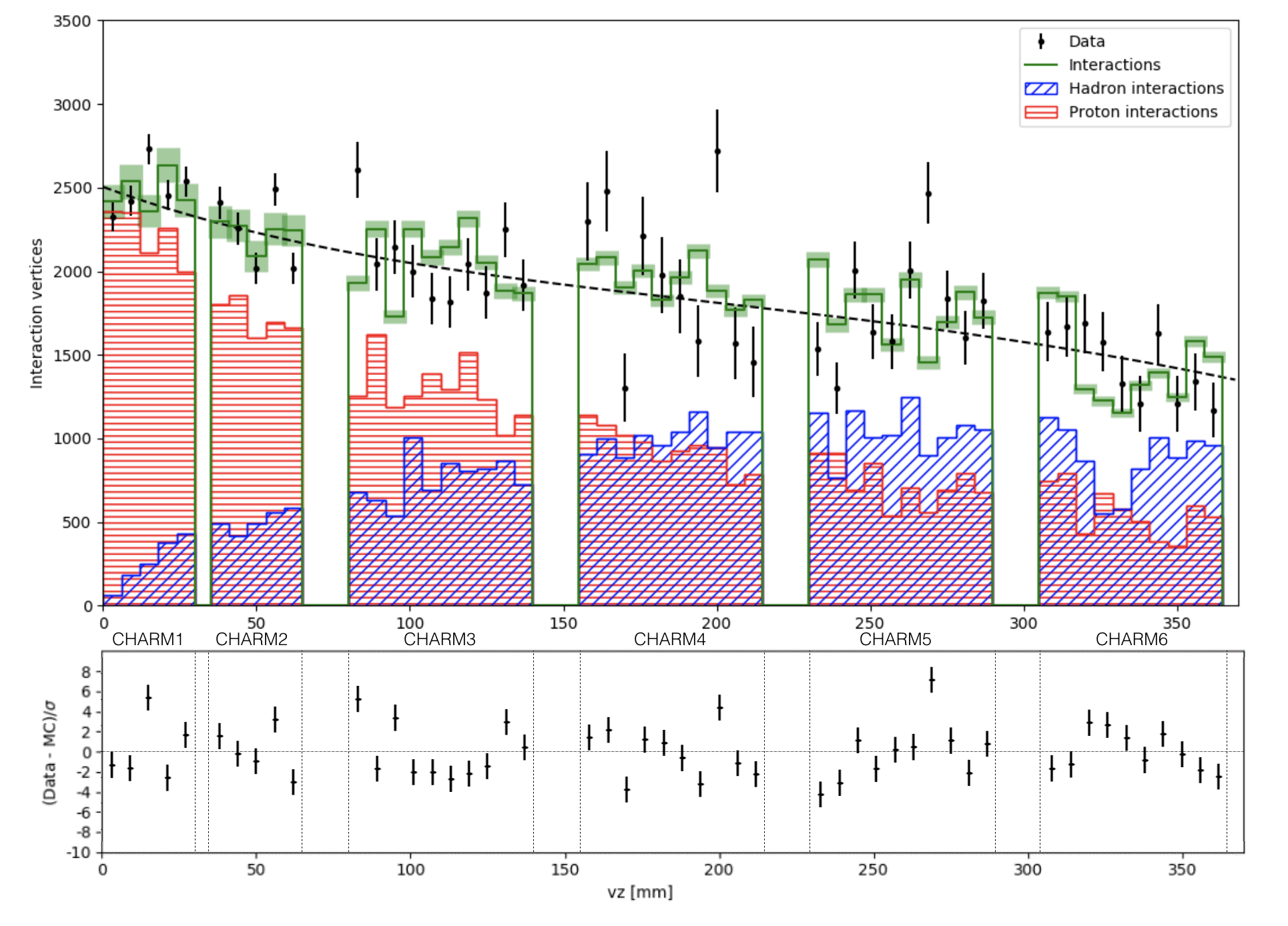}
\caption{Position distribution of interaction-vertex candidates along the beam direction for data and Monte Carlo, merging results from the different configurations. Primary-proton and hadron-reinteraction components are shown in red and blue, respectively. The dashed line represents the fit to data points. Empty bins refer to regions where data points are not available since they correspond to marginal regions of consecutive target configurations where the vertex reconstruction is not possible due to geometrical acceptance}.\label{fig:vz_tot}
\end{figure*}

By adding data reconstructed in different runs and combining the six configurations it is possible to retrieve the overall distribution of interaction-vertex candidates in a $\sim$365 mm long emulsion/lead target.
The overall distribution is shown in Figure~\ref{fig:vz_tot} for data and simulation. Error bars on data points are obtained propagating the covariance matrix of the original histogram with the efficiency correction factor.
\begin{comment}
Error bars on data are obtained by propagating the covariance matrix of the original histogram $nu_i$, 	$V_{ij} = \delta_{ij} \nu_j$, with the efficiency correction:

\begin{equation}
	U_i = \frac{1}{\epsilon} \nu_i \frac{1}{\epsilon} = \frac{1}{\epsilon}^2 \nu_i 
\end{equation}
\end{comment}

The distribution shown in  Figure~\ref{fig:vz_tot} is made by the sum of two components: primary protons and hadron reinteractions.
While the primary-proton component follows an exponential distribution, hadron reinteractions can be parametrised as a second-order polynomial. 
A Chi-square fit was therefore performed on data points with an exponential function and a 2$^{\mathrm{nd}}$ degree polynomial. The area under the two curves results to be 58\% and 42\%, respectively.

The exponent of the exponential function provides an estimation of the proton interaction length in the emulsion/lead target of

\begin{displaymath}
\lambda_I^{\mathrm{meas}} = (182 ^{+19}_{-16} ) \mbox{ mm}.
\end{displaymath}

Errors are purely statistical. %Systematic effects that 

This result is compatible with expectations from the full simulation, that predicts an interaction length of $(175 \pm 5 )$\,mm.\\

%\begin{figure}[htbp]
%\centering
%\includegraphics[scale=0.7]{figs/interaction_length.png}
%\caption{Position distribution of primary proton interactions in the SHiP-charm emulsion/lead target, after the subtraction of the hadron re-interaction component.}\label{fig:interaction_length}
%\end{figure}%

%% file: sections/conclusions.tex
\section{Conclusions}

The analysis of the SHiP-charm emulsion data required the development of dedicated software and analysis tools to extract the signal from an unprecedented background rate. A good agreement between data and Monte Carlo expectations is found for the number of charged tracks defining the interaction vertex and the position of the vertex along the beam axis. These results prove the capability to reconstruct interaction vertices in a harsh environment. 

The development of a Monte Carlo simulation that accurately described reconstructed data and the application of multivariate analysis techniques allowed to extract the primary proton interaction component in a $\sim$365 mm long emulsion/lead target and to evaluate the effective interaction length. The results to be in good agreement with expectations.

%% file: sections/acknowledgments.tex
\section*{Acknowledgments}

 SHiP Collaboration acknowledges support from the following Funding Agencies: the National Research Foundation of Korea (with grant numbers of 2018R1A2B2007757, \linebreak
2018R1D1A3B07050649, 2018R1D1A1B07050701, \linebreak
 2017R1D1A1B03036042,
2017R1A6A3A01075752, \linebreak
 2016R1A2B4012302, and 2016R1A6A3A11930680);  \linebreak
 the Russian Foundation for Basic Research (RFBR, grant 17-02-00607) and the TAEK of Turkey.

This work is supported by a Marie Sklodowska-Curie Innovative Training Network Fellowship of the European Commissions Horizon 2020 Programme under contract number 765710 INSIGHTS.

We thank M. Al-Turany, F. Uhlig. S. Neubert and A. Gheata their assistance with FairRoot. We acknowledge G. Eulisse and P.A. Munkes for help with Alibuild.

The measurements reported in this paper would not have been possible without a significant financial contribution from CERN. In addition, several member institutes made large financial and in-kind contributions to the construction of the target and the spectrometer sub detectors, as well as providing expert manpower for commissioning, data taking and analysis. This help is gratefully acknowledged.

%% file: authorlist_15march2024.tex
%\documentclass{article}
%\usepackage[utf8]{inputenc}
%\begin{document}
\onecolumn
\noindent
\textbf{The SHiP Collaboration}
\\
%\centerline{\Large\bf The SHiP Collaboration}
%\vspace*{1mm}
%\begin{flushleft}
%-- 
%-- SHiP Authorlist as of 26 August 2023
%-- 
\noindent
C.~Ahdida$^{32}$,
A.~Akmete$^{10}$,
R.~Albanese$^{15,d,h}$,
F.~Alicante$^{15,d}$,
J.~Alt$^{7}$,
A.~Alexandrov$^{15,d}$,
S.~Aoki$^{18}$,
G.~Arduini$^{32}$,
J.J.~Back$^{42}$,
F.~Baaltasar~Dos~Santos$^{32}$
% ...
\vspace*{1cm}

{\footnotesize \it

% ...
$^{7}$Physikalisches Institut, Universit\"{a}t Freiburg, Freiburg, Germany\\
% ...
$^{10}$Institut f\"{u}r Physik and PRISMA Cluster of Excellence, Johannes Gutenberg Universit\"{a}t Mainz, Mainz, Germany\\
% ...
$^{15}$Sezione INFN di Napoli, Napoli, Italy\\
% ...
$^{18}$Kobe University, Kobe, Japan\\
% ...
$^{32}$European Organization for Nuclear Research (CERN), Geneva, Switzerland\\
% ...
$^{42}$University of Warwick, Warwick, United Kingdom\\
% ...
$^{d}$Universit\`{a} di Napoli ``Federico II``, Napoli, Italy\\
% ...
$^{h}$Consorzio CREATE, Napoli, Italy\\
% ...
}

Following the resolution of the CERN Council, Russian groups participating in the
SHiP project are not included in the author list of this document. We acknowledge the
contribution of the Russian scientists, in particular A.~Anokhina, E.~Atkin, N.~Azorskiy, A.~Bagulya, A.~Baranov, A.Y.~Berdnikov, Y.A.~Berdnikov, A.~Chepurnov, M.~Chernyavskiy, L.~Dedenko, P.~Dergachev, V.~Dmitrenko, A.~Dolmatov, S.~Donskov, T.~Enik, A.~Etenko, O.~Fedin, K.~Filippov, G.~Gavrilov, V.~Golovtsov, D.~Golubkov, D.~Gorbunov, S.~Gorbunov, M.~Gorshenkov, V.~Grachev, V.~Grichine, N.~Gruzinskii, Yu.~Guz, M.~Hushchyn, D.~Karpenkov, M.~Khabibullin, E.~Khalikov, G.~Khaustov, A.~Khotyantsev, V.~Kim, A.~Kolesnikov, V.~Kolosov, N.~Konovalova, I.~Korol'ko, I.~Krasilnikova, Y.~Kudenko, E.~Kurbatov, P.~Kurbatov, V.~Kurochka, E.~Kuznetsova, V.~Maleev, A.~Malinin, A.K.~Managadze, A.~Mefodev, Yu.~Mikhaylov, O.~Mineev, S.~Movchan, S.~Nasybulin, B.~Obinyakov, N.~Okateva, D.~Pereyma, A.~Petrov, D.~Podgrudkov, V.~Poliakov, N.~Polukhina, M.~Prokudin, F.~Ratnikov, T.~Roganova, V.~Samoylenko, V.~Samsonov, E.S.~Savchenko, A.~Shakin, P.~Shatalov, T.~Shchedrina, V.~Shevchenko, A.~Shustov, M.~Skorokhvatov, S.~Smirnov, N.~Starkov, P.~Teterin, S.~Than~Naing, S.~Ulin, E.~Ursov, A.~Ustyuzhanin, Z.~Uteshev, L.~Uvarov, I.~Vidulin, K.~Vlasik, A.~Volkov, R.~Voronkov, Yu.~Zaytsev, A.~Zelenov.

% \end{flushleft}

% \end{document}